\documentclass[superscriptaddress,10pt,aps,pra,twocolumn,longbibliography]{revtex4-2}
\usepackage{float}
\floatstyle{boxed}
\usepackage{array}
\usepackage{graphicx}
\usepackage{amsbsy}
\usepackage[utf8x]{inputenc}
\usepackage{epstopdf}
\usepackage{amsmath,amssymb,amsfonts,amsthm}
\usepackage{indentfirst}
\usepackage{soul}
\usepackage[T1]{fontenc}
\usepackage[dvipsnames]{xcolor}
\usepackage{url}
\usepackage[colorlinks]{hyperref}
\usepackage{array}
\usepackage{bbm}
\usepackage{dsfont}

\hypersetup{
    plainpages=true,
    breaklinks=true,
    hypertexnames=false,
    pageanchor=true,
    colorlinks=true,
    linkcolor={blue},
    citecolor={red},
    urlcolor={blue},
    anchorcolor={black}
}

\newcommand{\figref}[1]{\mbox{Fig.~\ref{#1}}}

\renewcommand{\eqref}[1]{\mbox{Eq.~(\ref{#1})}}

\newcommand{\be}{\begin{equation}}
\newcommand{\ee}{\end{equation}}
\newcommand{\bea}{\begin{eqnarray}}
\newcommand{\eea}{\end{eqnarray}}

\newcommand{\De}{{D_\theta}}
\newcommand{\PH}{H^{{\rm pH}}}
\newcommand{\PV}{\psi^{\rm pH}}
\newcommand{\HH}{H^{{\rm H}}}

\begin{document}

\title{Achieving the Quantum Fisher Information Bound in Pseudo-Hermitian Sensors}
\author{Ievgen I. Arkhipov}
\email{ievgen.arkhipov@upol.cz}
\affiliation{Joint Laboratory of
Optics of Palack\'y University and Institute of Physics of CAS,
Faculty of Science, Palack\'y University, 17. listopadu 12, 771 46
Olomouc, Czech Republic}

\author{Franco Nori}
\affiliation{Quantum Information Physics
Theory Research Team, Quantum Computing Center, RIKEN, Wakoshi, Saitama, 351-0198, Japan} \affiliation{Physics Department, The University of Michigan, Ann Arbor, Michigan 48109-1040, USA}

\author{\c{S}ahin K. \"Ozdemir}
\affiliation{Department of Electrical and Computer Engineering, Saint Louis University, St.~Louis, Missouri 63103, USA}

\begin{abstract}
Non-Hermitian systems have attracted considerable interest over the last few decades due to their unique spectral and dynamical properties not encountered in Hermitian counterparts. An intensely debated question is whether non-Hermitian systems, described by pseudo-Hermitian Hamiltonians with real spectra, can offer enhanced sensitivity for parameter estimation when they are operated at or close to exceptional points. However, much of the current analysis and conclusions are based on mathematical formalism developed for Hermitian
quantum systems, which is questionable when applied to pseudo-Hermitian Hamiltonians, whose Hilbert space metric is intrinsically  parameter dependent.
Here, we develop a covariant formulation of quantum Fisher information (QFI) defined on the deformed Hilbert space of pseudo-Hermitian Hamiltonians. This covariant framework ensures the preservation of the state norm and enables a consistent treatment of parameter sensitivity.  We further show that the covariant QFI of  pseudo-Hermitian systems is dual to the ordinary QFI of corresponding Hermitian systems. 
Importantly, this correspondence naturally imposes an upper bound on the covariant QFI and allows one to identify  optimal projections which saturate the corresponding classical Fisher information to this ultimate limit. The developed framework also enables to set the criteria under which pseudo-Hermitian sensors can exhibit an advantage over their Hermitian counterparts of the same dimensionality.

\end{abstract}

\date{\today}

\maketitle

{\it Introduction.}---Non-Hermitian systems described by pseudo-Hermitian Hamiltonians possessing real spectra, have recently attracted considerable interest in part due to the possibility to utilize them for enhanced quantum sensing. Specifically, these systems exhibit so-called exceptional points (EPs) in their spectrum, where both eigenvalues and eigenvectors of the Hamiltonian coalesce~\cite{KatoBOOK}.
Near these EPs, the eigenvalue response to perturbations becomes non-analytic, scaling as a fractional power of the perturbation, which has been suggested as a mechanism for enhanced parameter sensitivity in quantum metrology~\cite{Wiersig2014,Wiersig2016,Ozdemir2019}.

Several studies have demonstrated that quantum sensors operating near EPs can exhibit significant amplification of signal response, potentially outperforming their Hermitian counterparts in both classical and quantum regimes~\cite{Liu2016,Chen2017,Hodaei2017,Lau2018,Lai2019,Bao2021,Bao2024,Kononchuk2022,Yu2020,Zhang2019,Peters2022,Yu2024,Wu2025_sens,Naikoo2025}. In parallel, other theoretical and experimental proposals suggested utilizing pseudo-Hermiticity away from EPs for quantum sensing~\cite{Chu2020,Xiao2024}.  However, this apparent sensing supremacy has also been scrutinized. In particular, concerns have been raised about whether the increased sensitivity near EPs translates into a genuine advantage when quantum noise and other measurement imperfections are properly taken into account~\cite{Zhang2018_peter,Langbein2018,Wang2020,Chen2019,Loughlin2024,Naikoo2023,Peng2024,Zheng2025}. 

Recent studies have also explored the fundamental limits of quantum sensing in pseudo-Hermitian systems through the lens of quantum Fisher information (QFI)~\cite{Ma2011}. In particular, it has been argued that pseudo-Hermitian systems offer no intrinsic advantage in parameter estimation over their Hermitian counterparts when the optimal measurement strategy is employed~\cite{Ding2023,Zeng2025}. These conclusions, however, rely on specific assumptions; most notably, the use of the dilation method, which introduces additional resources and encodes the system’s parameter dependence solely in the reduced non-Hermitian Hamiltonian. This also raises important questions about the generality of such no-go results. Furthermore, it remains unclear whether the conventional QFI formalism is directly applicable in pseudo-Hermitian settings, as naive application may lead to misleading or inconsistent results due to the nontrivial geometry of the underlying Hilbert space of pseudo-Hermitian Hamiltonians. 
Initial efforts to address this issue were undertaken in Ref.~\cite{Yu2024,naikoo2025metric}, where the QFI was reformulated using an appropriate metric on the deformed Hilbert space. Yet, that approach does not account for intricate system dynamics on such non-flat space.

In this Letter, we develop a general covariant quantum Fisher information (CQFI) framework for states governed by pseudo-Hermitian Hamiltonians. Introducing a covariant state derivative, our formulation faithfully captures the deformed Hilbert-space geometry while preserving state normalization. We show that the CQFI maps onto the standard QFI (SQFI) of an associated Hermitian system and, thus, reveal the duality between the two.  This correspondence naturally sets an upper bound on the CQFI and enables one to determine the optimal measurement projections required to saturate the corresponding classical Fisher information to this ultimate limit.
Crucially, our framework establishes criteria under which pseudo-Hermitian sensors can surpass Hermitian counterparts of the same Hilbert-space dimension. This dimension-preserving comparison contrasts with dilation-based approaches, where pseudo-Hermitian systems are benchmarked against higher-dimensional Hermitian embeddings~\cite{Ding2023}.

{\it Preliminaries on the Geometry of Pseudo-Hermitian Hilbert Spaces.}---We begin by outlining key geometric concepts of the metric-modified Hilbert space induced by pseudo-Hermitian Hamiltonians~\footnote{Pseudo-Hermitian Hamiltonians can be physically realized through several approaches. For instance, they can be implemented via the Naimark dilation method, where a pseudo-Hermitian system can be embedded into a higher-dimensional Hermitian system~\cite{Gunter2008}. Alternatively, one can realize them through the postselection procedure, where by properly monitoring the environment of an open quantum system, one can effectively induce the continuous non-unitary dynamics~\cite{Minganti2020,arkhipov2025b}}.
As in Hermitian quantum mechanics, the dynamics of a state vector \( |\psi^{\mathrm{pH}}(t)\rangle \) in a pseudo-Hermitian system is governed by the pseudo-Hermitian Hamiltonian $\PH$ in the Schr\"odinger equation (by setting $\hbar=1)$:
$
    i\partial_t |\psi^{\mathrm{pH}}(t)\rangle = H^{\mathrm{pH}} |\psi^{\mathrm{pH}}(t)\rangle.
$

To ensure that the state evolution is consistent with probability conservation, as in standard quantum mechanics, one must equip the Hilbert space with a positive-definite Hermitian metric operator $\eta$. This metric defines a modified inner product,
$
    \langle \psi | \phi \rangle_\eta = \langle \psi | \eta | \phi \rangle,
$
which induces a Riemannian-like geometric structure on the $\eta$-defomred Hilbert space, such that the norm \( \langle \psi^{\mathrm{pH}}(t) | \eta | \psi^{\mathrm{pH}}(t) \rangle \) is preserved in time.
For pseudo-Hermitian Hamiltonians with real spectrum,  the metric obeys the following condition~\cite{MOSTAFAZADEH_2010}
$
    (H^{\mathrm{pH}})^\dagger \eta = \eta H^{\mathrm{pH}}.
$
 In this case, the Hamiltonian $\PH$ generates a unitary-like evolution in the metric-modified Hilbert space, even though it is not Hermitian.

An important consequence of the pseudo-Hermiticity is that any pseudo-Hermitian Hamiltonian \( H^{\mathrm{pH}} \) can be mapped  to a Hermitian Hamiltonian \( H^{\mathrm{H}} \) through a similarity transformation~\cite{Ju2022}. Explicitly, since the metric operator \( \eta \) can always be written as
    $\eta = S^\dagger S$,
for some invertible operator \( S \), then the associated Hermitian Hamiltonian (up to an arbitrary unitary) can be defined as follows~\footnote{Note that the associated Hermitian Hamiltonian $\HH$ is not unique given a metric $\eta$, since by redefining $S\to TS$, where $T$ is some unitary, the metric remains invariant, though the Hermitian Hamiltonian is transformed as $\HH\to T\HH T^\dagger$. Since this unitary freedom corresponds to a gauge choice and therefore does not modify the underlying physics, one may, without loss of generality, choose 
$T$ to be parameter independent.}
$H^{\mathrm{H}} = S H^{\mathrm{pH}} S^{-1}$, $(\HH)^\dagger=\HH$.
This relation shows that pseudo-Hermitian quantum mechanics can be unitarily equivalent (in the \( \eta \)-inner product sense) to standard Hermitian quantum mechanics. Consequently, the corresponding state vectors in the flat and $\eta$-modified Hilbert spaces are related by
\begin{equation}\label{psih_psip}
    |\psi^{\mathrm{H}}\rangle = S |\psi^{\mathrm{pH}}\rangle.
\end{equation}
Although, in certain cases, the operator $S$ can be directly determined from the shape of the $\PH$, in general, however, it can be inferred from the metric~\footnote{See Supplementary Material}.

{\it Covariant derivative of states on non-flat Hilbert space.}---In order to consistently describe the vector transport on the curved space, one can borrow the concepts of connection and covariant derivative from differential geometry, which allows one to properly compare vectors at different points on a given curved manifold~\cite{NakaharaBook}. This is exactly the case for the $\eta$-deformed Hilbert space. Indeed, when the Hamiltonian $\PH$ depends on a continuous parameter $\theta$, the variation in \( \theta \) can induce an additional geometric structure in its non-flat Hilbert space. That is, the parameter space can be viewed as a base manifold over which the Hilbert space forms a vector bundle. 
Consequently, one can define a connection $\Gamma_\theta$ associated with the $\theta$-direction in order to enable a proper covariant differentiation with respect to this parameter. This newly induced connection  ensures that a covariant state derivative respects the Hilbert space curvature induced by the metric \( \eta(\theta) \).  The explicit form of such a connection can read~\cite{Note3}:
  $ \Gamma_\theta = S^{-1} \partial_\theta S$. Moreover, when the eigenvectors of the metric $\eta$ do not depend on the parameter $\theta$, the connection can be alternatively written as $\Gamma_\theta = \eta^{-1} \partial_\theta \eta/2$~\cite{Note3}.
\color{black}

The associated covariant derivative along $\theta$ acting on the state vector $|\psi\rangle$ can then be defined as~\cite{NakaharaBook,Note3}
\begin{equation}\label{D}
    D_\theta |\psi\rangle = \partial_\theta |\psi\rangle + \Gamma_\theta |\psi\rangle.
\end{equation}
Clearly, when the Hilbert space is flat, i.e., $\eta=I$, the covariant derivative  is reduced to the ordinary derivative $\De=\partial_\theta$, since $\Gamma_\theta=0$ in this case~\footnote{The covariant derivative with the metric-compatible connection in \eqref{D}  satisfies the following relation~\cite{NakaharaBook}: $\partial_\theta\langle\psi|\psi\rangle_\eta=\langle D_\theta\psi|\psi\rangle_\eta+\langle\psi|D_\theta\psi\rangle_\eta$. In particular, when $\eta=I$, implying $\Gamma=0$, the covariant derivative is reduced to the standard derivative $D_\theta=\partial_\theta$. As a result, the derivative of the scalar $\partial_\theta\langle\psi|\psi\rangle$ simplifies to the ordinary derivative of the inner product of the two state-vectors on the standard flat Hilbert space.}.

{\it Covariant Quantum Fisher Information.}---The accuracy with which one can estimate a parameter 
$\theta$ encoded in a quantum state is fundamentally constrained by the quantum Cramér-Rao inequality~\cite{Helstrom1967}. According to this bound, the standard deviation 
$\Delta\theta$ of any unbiased estimator is limited from below by $\Delta\theta\geq1/\sqrt{m F_\theta}$, where 
$m$ denotes the number of independent measurements and $F_\theta$ represents the QFI associated with the state.

The SQFI for a pure state $|\psi^{\rm H}(\theta)\rangle$ of a Hermitian Hamiltonian $\HH$, parameterized by a real parameter $\theta$, is related to the Fubini–Study metric on the projective Hilbert space (the manifold of rays) \cite{Braunstein1994}:
\begin{equation}\label{SQFI}
F^{\rm H}(\theta) = 4 ||\partial_\theta \psi^{\rm H}_\perp ||^2,
\end{equation}
where we defined the vector's norm as $|| \cdot ||^2 = \langle \cdot | \cdot \rangle$, and
$
|\cdot_\perp\rangle = |\cdot\rangle - |\psi^{\rm H}\rangle \langle \psi^{\rm H} | \cdot \rangle
$
is the projection of a vector $|\cdot\rangle$ onto the subspace orthogonal to the state $|\psi^{\rm H}\rangle$.

In pseudo-Hermitian systems,  one has to introduce the covariant derivative instead, which faithfully accounts for the change of a state $|\psi^{\rm pH}(\theta)\rangle$ in a deformed Hilbert space.
In other words, if one naively applies the SQFI to pure states of pseudo-Hermitian systems, it would ignore the underlying geometric contribution to the QFI. 
To account for the geometry, one must generalize a SQFI to CQFI, defined as
\begin{equation} \label{CFQI}
F_c^{\rm pH}(\theta) = 4 \left|\left| D_\theta \psi^{\rm pH}_\perp \right|\right|^2_\eta,
\end{equation}
where the $\eta$-norm is given by $||\cdot \|_\eta^2 = \langle \cdot | \eta | \cdot \rangle$, and the perpendicular component is again projected as $|\cdot_\perp\rangle = (I - P_\psi) |\cdot\rangle$, with the properly redefined projector $P_\psi = |\psi^{\rm pH}\rangle_\eta \langle \psi^{\rm pH}|$.

{\it Duality between CFQI and SQFI.}---With the  covariant framework established above one finds a one-to-one correspondence between the SQFI of the Hermitian Hamiltonian and the CQFI of the pseudo-Hermitian system~\cite{Note3}:
\begin{equation}
 F^{\rm H}(\psi^H_\theta)=F^{\rm pH}_c(\PV_{\theta}).
\label{equiv}
\end{equation}
This duality establishes CQFI as the natural extension of quantum Fisher information to pseudo-Hermitian systems and constitutes the central result of our work.

The equivalence in \eqref{equiv} entails an important consequence. Namely, the upper bound on the  CQFI is identical to that on the SQFI. However, the upper bound of the SQFI is known, and it equals to~\cite{Giovanetti2006}
$
    F^{\rm H}_{\rm max}=(\lambda_{\rm max}-\lambda_{\rm min})^2,
$
where the $\lambda_{\rm max}\ (\lambda_{\rm min})$ is the maximal (minimal) eigenvalue of the Hermitian operator $h(\theta)=iV\partial_\theta V^{\dagger}$, where $V=\exp(-i\HH t)$, and therefore one has 
\begin{equation}\label{CQFIup}
    F^{\rm pH}_c(\PV_{\theta})\leq F^{\rm pH}_{\rm max}(\theta)= F^{\rm H}_{\rm max}(\theta).
\end{equation}
 For small values of $t\ll1$, the Hermitian operator is substantially simplified and takes the form $h(\theta)\approx t\partial_\theta \HH$~\cite{Wilcox1967}. In this case, the upper bound on the CQFI is straightforwardly found from the spectrum width of $\partial_\theta\HH$. For arbitrarily large $t>1$, the explicit formula for $h(\theta)$ 
 has been given in Ref.~\cite{Pang2014}.
 Importantly, the expression for the upper bound on the CQFI in \eqref{CQFIup} immediately gives the states $|\PV\rangle$ which maximize the CQFI, and which explicitly read $|\PV_{\rm max}\rangle\equiv S^{-1}\left[|\lambda_{\rm max}\rangle+e^{i\phi}|\lambda_{\rm min}\rangle\right]$, where $|\lambda\rangle$ is an eigenstate of the operator $h(\theta)$, and $\phi$ is an arbitrary phase. 

At first glance, the revealed duality in~\eqref{equiv} suggests that pseudo-Hermitian systems offer no fundamental sensing advantage over their Hermitian counterparts: under optimal measurements, sensitivities coincide. These preliminary conclusions agree with Refs.~\cite{Ding2023,Zeng2025}, which, however, rely solely on the quantum dilation framework, 
thus necessitating additional resources and assumptions.

Nevertheless, and most crucially, in the linear sensing regime pseudo-Hermitian systems can still exhibit enhanced performance. This is because linear perturbations in the pseudo-Hermitian frame may map to nonlinear perturbations in the Hermitian frame via the similarity transformation $S$, thus effectively amplifying sensitivity. It is well established that, when the parameter under interest couples to the Hermitian system nonlinearly, it results in enhanced sensing precision~\cite{Boixo2008}. We further substantiate these conclusions below by analyzing two minimal paradigmatic models.

{\it Optimal projections saturating the CQFI bound.}--- 
In a real experiment, however, any measurement on a quantum system yields a classical Fisher information (CsFI), determined by the measured probabilities $p(x_\theta|\theta)$, which reaches the QFI only when optimal state projections are applied~\cite{Giovanetti2006,Giovannetti2011}. These optimal projections can be determined from the symmetric logarithmic derivative (SLD) operator ${L}$, defined in flat Hilbert space for a state $\rho_\theta$ as $\dot{\rho}_\theta=(L\rho+\rho L)/2$~\cite{Helstrom1976,Holevo1982}.
The significance of ${L}$ lies in the fact that its variance equals the QFI,
$F_Q=\langle L^{2}\rangle$, since $\langle L \rangle=0$,
so the eigenvectors of the SLD are precisely the optimal states that saturate the CsFI to the QFI~\cite{Braunstein1994,Paris2009}. For pure states the SLD operator attains the form $L=2(|\partial_\theta\psi\rangle\langle\psi|+|\psi\rangle\langle\partial_\theta\psi|)$.

In this regard, we can promote the standard SLD operator, defined on the flat Hilbert space, to a covariant SLD operator on the $\eta$-modified Hilbert space of pseudo-Hermitian systems. Indeed, the correspondence in~\eqref{equiv} implies that the variances of the ordinary $L$ and covariant $L_c$ SLD operators, averaged over the corresponding states in \eqref{psih_psip}, must be dual to each other, i.e.,
$\langle L^{2}\rangle=\langle L_c^{2}\rangle_\eta$, where the pseudo-Hermitian $L_c$ operator reads 
\begin{eqnarray}\label{L}
    L_c=2(|D_\theta\psi\rangle\langle\psi|_\eta+|\psi\rangle\langle D_\theta\psi|_\eta), \quad L_c^{\dagger}=\eta L_c\eta^{-1}.
\end{eqnarray}
The correspondence between these two SLD operators immediately provides a prescription for finding the optimal projection in pseudo-Hermitian sensors that enables the ultimate precision in parameter estimation. Indeed, knowing the state $|\psi^{\rm pH}_{\rm max}\rangle$, which maximizes the CQFI, allows one to find the corresponding optimal projectors from the eigenvectors  $|\nu_i\rangle$ of $L_c$ in \eqref{L}.

Therefore, realizing measurements via the corresponding projectors $\Pi_i = |\nu_i\rangle\langle\nu_i|_\eta$ enables one to saturate the covariant CsFI, 
\begin{eqnarray}\label{Fcl}
    F_{\rm cl}^{\rm pH}(\theta)=\sum{(\partial_\theta p_i)^2}/{p_i}, 
    \  p_{i}(\theta)=\langle\psi^{\rm pH}_{\rm max}|\Pi_i|\psi^{\rm pH}_{\rm max}\rangle_\eta,
\end{eqnarray}
to the upper bound of the CQFI, i.e., $F_{\rm cl}^{\rm pH}(\theta)=F_{\rm max}^{\rm pH}(\theta)$.  

Needless to say, performing projections with respect to the $\eta$-modified inner product on the associated Hilbert space can be experimentally challenging. However, recent experiments have demonstrated the feasibility of metric-dependent projections in various non-Hermitian quantum platforms~\cite{Zhang2025,Wang2025,Wang2024}, thus indicating that achieving {\it `covariant'} ultimate precision in pseudo-Hermitian quantum sensors is now within experimental reach.
\color{black}

{\it Nonreciprocal systems. Example 1.a.}---
Now we examine conditions when quantum sensing can or cannot exhibit enhancement in the pseudo-Hermitian systems in comparison to their Hermitian counterparts.
We first analyze a specific setup analogous to that in Refs.~\cite{Chu2020, Xiao2024, Zeng2025}, where the Hamiltonian is parameterized by a multiplicative parameter $\theta$, namely
\begin{equation}\label{He1}
    \PH(\theta)=\theta\begin{pmatrix}
        0& \delta^{-1} \\
        \delta & 0
    \end{pmatrix}, \quad \delta\neq 0
\end{equation}
The form of the metric operator is readily found $\eta={\rm diag}[1,\delta^{-2}]$~\cite{Note3}, which does not depend on $\theta$.
The associated $\HH$ gets the form~\cite{Note3}:
  $  \HH(\theta)=\theta\begin{pmatrix}
        0& 1 \\
        1 & 0
    \end{pmatrix}$.

Comparing both Hamiltonians, one comes to the conclusion that for this type of parameterization, the $\PH$ {\it does not} provide any advantage in enhancing the QFI, since the multiplicative nature of the parameterization remains in the associated Hermitian Hamiltonian $\HH$. Specifically, for $t\ll 1$, the upper bound on the CQFI for this type of pseudo-Hermitian Hamiltonian reads
\begin{equation}
    F^{\rm pH}_c(\theta)\leq 4t^2,
\end{equation}
which does not depend on $\theta$.

\paragraph{Example 1.b.---}
On the other hand, by choosing a parameterization as follows
\begin{equation}\label{He3}
    \PH(\theta)=\begin{pmatrix}
        0& \delta^{-1}+\theta \\
        \delta+\theta & 0
    \end{pmatrix}, \quad \delta\neq 0,
\end{equation} 
one can achieve enhancement in quantum sensing. This kind of nonreciprocal sensor can be implemented in quantum photonic settings~\cite{Soleymani2022,Zhang2025b}.
The associated $\HH$ has the perturbed off-diagonal terms $\sqrt{(\delta+\theta)(\delta^{-1}+\theta)}$~\cite{Note3}, which are nonlinear in $\theta$, in contrast to linear perturbations in \eqref{He3}. 
This means that for the same fixed measurement protocol based on a linear disturbance, one can achieve better performance in the QFI in the pseudo-Hermitian system governed by $\PH$ in \eqref{He3}. 

\begin{figure}[t!]
    \includegraphics[width=0.45\textwidth]{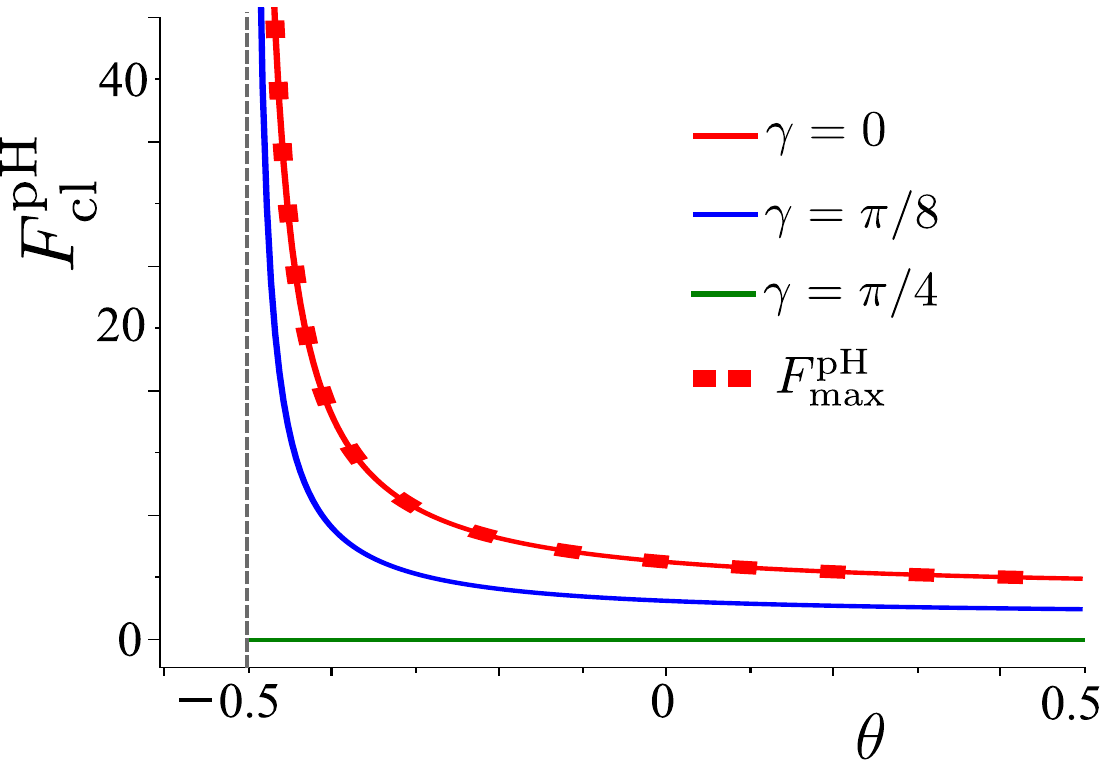}
    \caption{
   {Covariant classical Fisher information obtained from different measurement projectors
in the nonreciprocal system of Example~1.b. 
The parameter $\gamma$ controls the deviation of the measurement basis 
$|\chi(\gamma)\rangle$ from the optimal axis defined by the eigenbasis of the covariant SLD operator (see the main text for details).
The upper bound of the covariant quantum Fisher information for each case is shown by a red dotted curve. The vertical dash gray line corresponds to the value of $\theta=\theta_{\rm EP}=-0.5$ [arb. units], when the pseudo-Hermitian sensor is at the exceptional point.}}
    \label{fig1}
\end{figure}

Again, assuming for simplicity $t\ll 1$, the CQFI for this type of parameterization is bounded as follows
\begin{equation}\label{CQFIup2}
     F^{\rm pH}_c(\theta)\leq F^{\rm pH}_{\rm max}(\theta)= \frac{(\delta^2+2\theta\delta+1)^2}{\delta(\delta+\theta)(\delta\theta+1)}t^2.
\end{equation}
Clearly, the CQFI bound $F^{\rm pH}_{\rm max}(\theta)$ diverges as the system approaches the EP, which corresponds to the zeros in the denominator of \eqref{CQFIup2} and is associated with the Jordan form of the pseudo-Hermitian Hamiltonian in \eqref{He3}  [see also \figref{fig1}].  Consequently, initializing the system near the EP can enhance sensitivity in parameter estimation.  Importantly, the CQFI faithfully captures the phase transition occurring at the EP, signaling spontaneous pseudo-Hermiticity symmetry breaking. This observation is in line with Hermitian systems, where the SQFI is utilized as a witness for quantum phase transitions~\cite{Wang_2014,Yin2019,Guan2021,Montenegro2025}. 
It is also instructive to compare these results with those obtained within the flat Hilbert space formalism which leads to erroneous conclusions~\cite{Note3}.

Next we find the optimal projections, i.e., eigenstates of the covariant SLD operator in~\eqref{L}, which saturate the corresponding CsFI to the CQFI bound given in \eqref{CQFIup2}.
As the initial state we choose the vacuum $|0\rangle$. The time-evolving state is then 
$|\psi_t\rangle = e^{-i\PH t}|0\rangle$.  
In the basis $\{|0\rangle,|1\rangle\}$, the $\eta$-normalized state takes the form
\begin{equation}\label{0t}
    |\psi_t\rangle = \cos(\sqrt{k_1k_2}\,t)\,|0\rangle 
    - i\sqrt{\frac{k_2}{k_1}}\,\sin(\sqrt{k_1k_2}\,t)\,|1\rangle,
\end{equation}
which satisfies $\langle \psi_t|\eta|\psi_t\rangle=1$, with $\eta$ given in~\cite{Note3}. Here we denote  
$k_1=\delta^{-1}+\theta$ and $k_2=\delta+\theta$.  
A straightforward calculation shows that the CQFI of this state attains its upper bound $F^{\rm pH}_{\rm max}(\theta)$ in~\eqref{CQFIup2}, diverging when approaching the EP.

After some algebra, one finds that the eigenstates $|\nu_\pm\rangle$ of the $L_c$ operator take the form~\cite{Note3}  
\begin{equation}\label{Pi}
    |\nu_{\pm}\rangle = (|\psi_t\rangle \pm |u_t\rangle)/{\sqrt{2}},
\end{equation}
where
$|u_t\rangle = \sin(\sqrt{k_1k_2}\,t)\,|0\rangle 
    + i\sqrt{{k_2}/{k_1}}\,\cos(\sqrt{k_1k_2}\,t)\,|1\rangle$,
satisfying $\langle\psi_t|u_t\rangle_\eta=0$, and $\langle\nu_{i}|\nu_j\rangle_\eta=\delta_{ij}$. The optimal projectors then read $\Pi_\pm(\theta)=|\nu_\pm\rangle\langle\nu_\pm|_\eta$.

Evidently, the optimal projectors generally depend on the parameter $\theta$, which means that suitable measurement strategies are required to saturate the CsFI~\cite{Liu2024}. Such measurement approaches involve measuring the projections at the fixed basis at some reference points $x_0=(\delta_0,\theta_0,t_0)$, so that evaluating the corresponding probabilities 
$p_\pm(\theta)=\langle\psi_t(\theta)|\Pi_\pm(x_0)|\psi_t(\theta)\rangle_\eta$
 allows one to achieve the CQFI bound~\cite{Liu2024}. Remarkably, the projectors $\Pi_\pm(x_0)$ saturate the CsFI for any  fixed system parameters $x_0$, implying that these optimal projectors are universal and the corresponding nonreciprocal pseudo-Hermitian sensor is basis-robust~\footnote{{We note that in evaluating the probabilities $p_\pm(\theta)=\langle\psi_t(\theta)|\Pi_\pm(x_0)|\psi_t(\theta)\rangle_\eta$, the inner product involves states defined, in general, at different points in parameter space. Consequently, the states $|\nu(x_0)\rangle$ and $|\psi_t\rangle$ may correspond to distinct local metrics. This ambiguity can be resolved either by parallel transporting one state to the parameter location of the other, or by flattening the $\eta$-deformed Hilbert space. Here we adopt the latter approach and discuss it in the Supplementary Material.}}. 

In other words, even though the measurement outcomes $p_\pm(\theta,x_0)$ depend on the given choice of the basis $\Pi_\pm(x_0)$, this dependence vanishes in the expression for CsFI in~\eqref{Fcl}, 
thus leading to $F_{\rm cl}^{\rm pH}(\theta)=F^{\rm pH}_{\rm max}(\theta)$ (see Supplementary Material for details). For example, one can verify that the set $\{|0\rangle,|1\rangle\}$ is a particular choice of the eigenstates $|\nu_{\pm}\rangle$ in~\eqref{Pi} for certain $x_0$. Hence, by implementing sensing measurements in this frame already allows one to saturate the CsFI.

On the other hand, to assess the impact of {\it non-optimal} measurements on the CsFI, we consider projections onto 
\begin{equation}\label{nonopt1}
  \Pi(\gamma,\phi)=|\chi(\gamma,\phi)\rangle\langle\chi(\gamma,\phi)|_\eta, \quad \text{and} \quad {\mathds 1}-\Pi(\gamma,\phi),
\end{equation}
where $|\chi(\gamma,\phi)\rangle=\cos\gamma\,|\nu_+\rangle+e^{i\phi}\sin\gamma\,|\nu_-\rangle$ 
denotes a generic state rotated away from the optimal basis in~\eqref{Pi}. 
Substituting this form into~\eqref{Fcl} with $\phi=\pi/2$ and taking, without loss of generality, the projectors’ reference point $x_0$ to coincide with the parameter point of the probe state $|\psi_t\rangle$, the CsFI 
reduces to the `suppressed' form
\begin{equation}
    F_{\rm cl}^{\rm pH}(\theta,\gamma)=F^{\rm pH}_{\rm max}(\theta)\cos^22\gamma,
\end{equation}
as shown in Fig.~\ref{fig1}. The general case is discussed in~\cite{Note3}.

{In addition to nonreciprocal systems, another class of pseudo-Hermitian sensors, namely, $\cal PT$-symmetric systems, is often considered in the context of enhanced sensing. Accordingly, in the Appendix, we analyze these systems, deriving the upper bound of their CQFI and presenting the explicit form of the optimal projectors that saturate this limit.}

\color{black}
{\it Conclusions.}---In summary, we developed a covariant formulation of QFI for pure states evolving under pseudo-Hermitian Hamiltonians, properly accounting for the metric-modified geometry of the underlying Hilbert space. We also uncovered a duality between the standard and covariant QFI, which provides a natural upper bound for the latter; and explicit criteria for when pseudo-Hermitian systems can outperform Hermitian counterparts in quantum sensing. {Furthermore, this duality allows to define a covariant symmetric logarithmic derivative operator, whose eigenbasis determines the optimal projectors that saturate the covariant QFI bound. Recent advances in state measurements on metric-dependent Hilbert spaces~\cite{Zhang2025,Wang2025,Wang2024} make our results directly relevant for experiments, thus opening a path toward practically attaining ultimate precision in pseudo-Hermitian quantum sensors operating at or in close vicinity of EPs.}

\acknowledgements
I.A. acknowledges support from the Air Force Office of Scientific Research (AFOSR) Award No. FA8655-24-1-7376, 
and from the Grant Agency of the Czech Republic (Project No. 25-15775S).
F.N. is supported in part by:
the Japan Science and Technology Agency (JST)
[via the CREST Quantum Frontiers program Grant No. JPMJCR24I2,
the Quantum Leap Flagship Program (Q-LEAP), and the Moonshot R\&D Grant Number JPMJMS2061],
and the Office of Naval Research (ONR) Global (via Grant No. N62909-23-1-2074).
S.K.O.
acknowledges support of the 
AFOSR Multidisciplinary University Research Initiative (MURI)
Award on Programmable Systems with Non-Hermitian Quantum Dynamics
(Award No. FA9550-21-1-0202).


\appendix
\renewcommand{\theequation}{A\arabic{equation}}
\setcounter{equation}{0}
\onecolumngrid
\vspace{0.5cm}
\begin{center}
    \textbf{\large Appendix}
\end{center}
\vspace{0.5cm}
\twocolumngrid

{\it Appendix: Example 2. $\cal PT$-symmetric systems.}---Here we analyze another well-known type of pseudo-Hermitian systems, namely, $\cal PT$-symmetric systems. 

Specifically, we consider a $\cal PT$-symmetric Hamiltonian of the form
\begin{equation}\label{HPT}
    H^{\cal PT}=\begin{pmatrix}
        re^{i\tau} & s \\
        s & re^{-i\tau}
    \end{pmatrix}, \quad r,\tau,s\in\mathbb{R}.
\end{equation}
Such Hamiltonian can describe, e.g., a coupled optical dimer with loss and gain~\cite{Ozdemir2019}.
The $\cal PT$-symmetry condition is satisfied by identifying the parity operator with the Pauli matrix, i.e., ${\cal P}=\sigma_x$, and $\cal T$ with complex conjugation, such that $[H^{\cal PT},{\cal PT}]=0$. 
This Hamiltonian undergoes a spontaneous $\cal PT$-symmetry breaking at the EP defined with the relation $s_{\rm EP}=r\sin\tau$. Such that when $s>r\sin\tau$ the $H^{\cal PT}$ is in the exact $\cal PT$-phase, meaning that its spectrum is real. For $s<r\sin\tau$, on the other hand, the spectrum becomes complex, implying that the Hamiltonian is in the broken $\cal PT$-phase.

The metric operator $\eta$ for this $\cal PT$-symmetric Hamiltonian has been already calculated in Ref.~\cite{Ju2019}, which reads
\begin{equation}
    \eta=\dfrac{1}{\cos\alpha}\begin{pmatrix}
        1 & -i\sin\alpha \\
        i\sin\alpha & 1
    \end{pmatrix}, \quad \alpha=\arcsin{\left(\frac{r}{s}\sin\tau\right)}.
\end{equation}
By diagonalizing the metric $ \eta=U\Lambda U^\dagger$,
we find 
\begin{equation}
    \Lambda={\rm diag}\left[\sec\alpha+\sqrt{\sec^2\alpha-1},\sec\alpha-\sqrt{\sec^2\alpha-1}\right],
\end{equation}
and the corresponding unitary
\begin{eqnarray}
     U=\frac{1}{\sqrt{2}}\begin{pmatrix}
        -i & i \\
        1 & 1
    \end{pmatrix}.
\end{eqnarray}

 Knowing the metric eigenspace, one can straightforwardly find the associated Hermitian Hamiltonian $\HH$:
\begin{equation}\label{HH2}
    \HH=\begin{pmatrix}
        r\cos\tau & i\sqrt{s^2-r^2\sin^2\tau} \\
        -i\sqrt{s^2-r^2\sin^2\tau} & r\cos\tau
    \end{pmatrix}.
\end{equation}
Evidently, the Hamiltonian $\HH$ in \eqref{HH2} is Hermitian as long as the $H^{\cal PT}$ is in the unbroken $\cal PT$-phase. 

Having established the $\HH$, one can now determine the upper bound on the CQFI for a linearly perturbed pseudo-Hermitian system described by the perturbed Hamiltonian $H^{\cal PT}\to H^{\cal PT}+\theta\sigma_x$, where the symmetric coupling is modified by $\theta$ as $s\to s+\theta$. For sufficiently short evolution times $t\ll 1$, the upper bound attains the form
\begin{equation}\label{CQFIup3}
     F^{\cal PT}_c(\theta)\leq F^{{\cal PT}}_{\rm max}(\theta)=\dfrac{4(s+\theta)^2}{(s+\theta)^2-r^2\sin^2\tau}t^2.
\end{equation}
Again, the CQFI bound diverges when the system approaches the EP (the denominator becomes zero at the EP), implying the system undergoes a spectral phase transition corresponding to spontaneous $\cal PT$-symmetry breaking. Importantly, the linear perturbations in the $\cal PT$-symmetric systems are equivalent to the nonlinear perturbations in their Hermitian counterparts. Hence, the $\cal PT$-symmetric sensors can exhibit enhanced sensing supremacy over the Hermitian systems when the estimated parameter is encoded via the linear perturbations.

We now turn to the evaluation of the optimal projectors for this $\cal PT$-symmetric sensor, which saturates the CsFI to the CQFI bound given in~\eqref{CQFIup3}. 
As in Example~1.b in the main text, we focus on the time-evolving state starting from the vacuum, i.e., $ |\xi_t\rangle=\exp(-iH^{\cal PT}t)|0\rangle$, or more specifically 
\begin{equation}\label{xit}
    |\xi_t\rangle=S^{-1}(\cos\Omega t|0\rangle-\sin\Omega t|1\rangle),
\end{equation}
where $\Omega=\sqrt{(s+\theta)^2-r^2\sin^2\tau}$, and the matrix $S$, which is the similarity transformation between Hamiltonians in Eqs.~(\ref{HPT}) and (\ref{HH2}), explicitly reads as \begin{equation}\label{Spt}
    S=\sqrt{\Lambda}U^{\dagger}=\frac{1}{\sqrt{2}}\begin{pmatrix}
        i\kappa_+ & \kappa_+ \\
        -i\kappa_- & \kappa_-
    \end{pmatrix},
\end{equation}
with $\kappa_\pm=\sqrt{\sec\alpha\pm\tan\alpha}$, and $\alpha=\arcsin{\left(\frac{r}{s}\sin\tau\right)}$.

One can readily check that for the state $|\xi_t\rangle$ the CQFI attains its upper bound, i.e., 
 \begin{equation}
     F_c^{\cal PT}(\xi_t) = F^{\cal PT}_{\rm max}(\theta).
 \end{equation}
Accordingly, the optimal projectors follow directly from the eigenstates of the covariant SLD operator $L_c$ defined by the state $|\xi_t\rangle$ in~\eqref{L}, and which read:
\begin{equation}\label{mu}
    |\mu_{\pm}\rangle=S^{-1}(w_\pm|0\rangle\pm w_\mp|1\rangle)/\sqrt{2}, \quad \langle\mu_i|\mu_j\rangle_\eta=\delta_{ij}, 
\end{equation}
where $w_\pm=\cos\Omega t\pm\sin\Omega t$. 

\begin{figure}[t!]
    \includegraphics[width=0.48\textwidth]{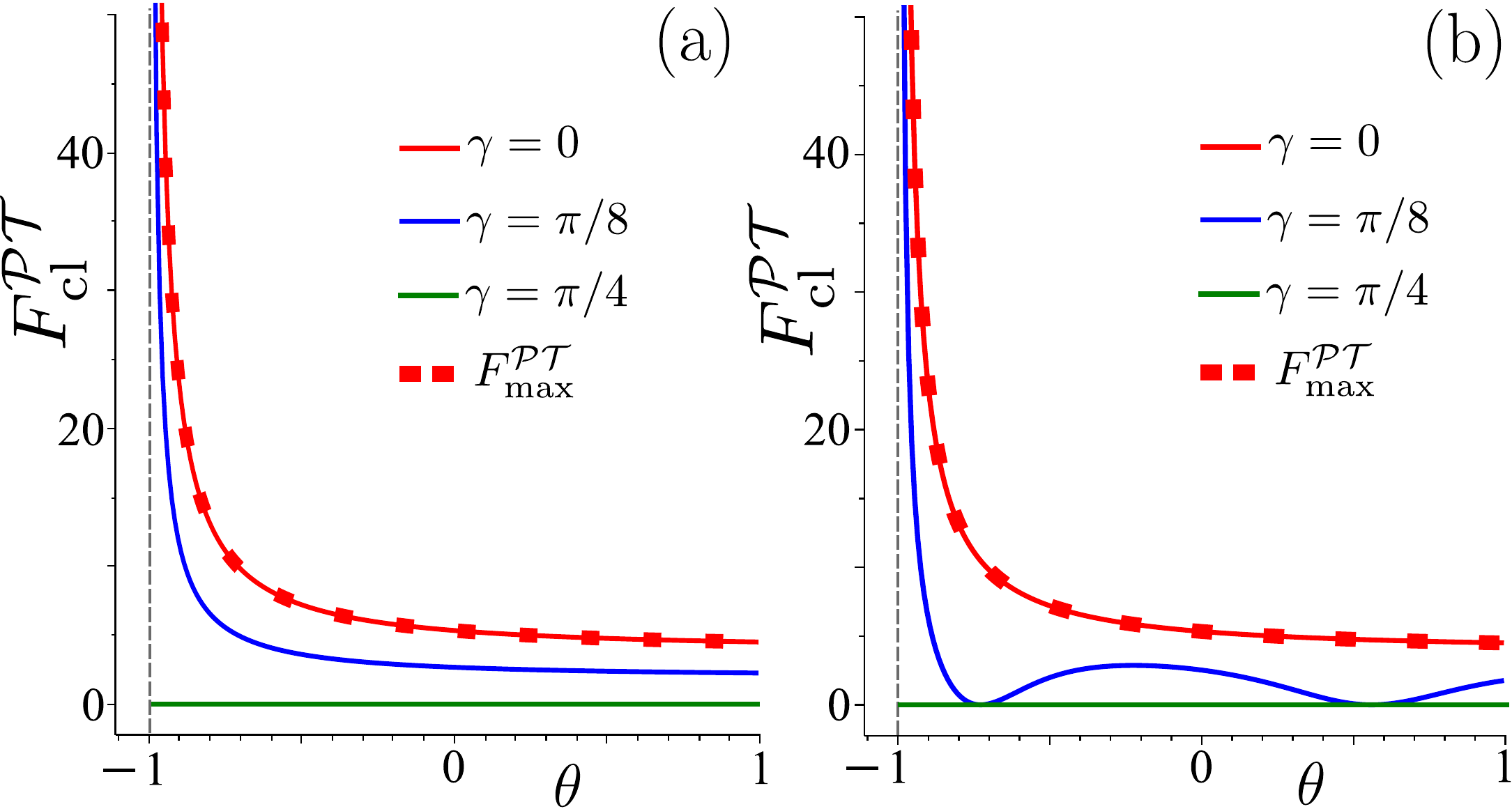}
    \caption{
  {Covariant classical Fisher information in the $\cal PT$-symmetric system, obtained from different measurement projectors: (a) when the reference point $x_0$ coincides with the parameter point of the probe state $|\xi_t\rangle$, and (b) when $x_0$ differs from the probe-state parameter point. 
The parameter $\gamma$ controls the deviation of the measurement basis 
$|\chi(\gamma)\rangle$ from the optimal axis defined by the eigenbasis of the covariant SLD operator.
The upper bound of the covariant quantum Fisher information for each case is shown by a red dotted curve. In both panels, the vertical dash gray line corresponds to the value of $\theta=\theta_{\rm EP}$, when the corresponding sensor is at the exceptional point.}}
    \label{fig2}
\end{figure}

Clearly, the eigenbasis of $L_c$ depends on the parameter $\theta$. Nonetheless, analogous to Example 1.b,  one  finds that by fixing the eigenbasis $|\mu_\pm(x_0)\rangle$ at an arbitrary reference point $x_0=(r_0,\tau_0,s_0,\theta_0,t_0)$ in the exact-$\cal PT$-phase, and by measuring the outcomes  
\begin{equation}\label{p0}
    p_\pm(\theta)=\langle\xi_t(\theta)|\Pi_\pm(x_0)|\xi_t(\theta)\rangle_\eta,
\end{equation}
with projectors  
\begin{equation}\label{Pi0}
    \Pi_\pm(x_0)=|\mu_\pm(x_0)\rangle\langle\mu_\pm(x_0)|_\eta,
\end{equation}
one can always saturate the corresponding CsFI to the CQFI bound, i.e.,  
$
F_{\rm cl}^{\cal PT}(\theta)=F^{\cal PT}_{\rm max}(\theta)
$. Indeed, by combining Eqs.~(\ref{p0}) and (\ref{Pi0}) one obtains
\begin{equation}\label{p01}
    p_{\pm}(\theta)=\frac{1\pm\sin2(\Omega_0 t_0-\Omega t)}{2}.
\end{equation}
Inserting now \eqref{p01} into \eqref{Fcl}, one arrives at
\begin{equation}
    F_{\rm cl}^{\cal PT}(\theta)=\dfrac{4(s+\theta)^2}{(s+\theta)^2-r^2\sin^2\tau}t^2=F^{\cal PT}_{\rm max}(\theta),
\end{equation}
thus confirming that the CsFI saturates the CQFI bound, independent of the chosen reference point $x_0$. Note that in the calculations above it is important to properly handle the metric-induced inner product for two states at different system parameter points, where the local metric may differ. We provide further details on this matter in Sec.~VII of the Supplementary Material.

To examine the effect of non-optimal measurements in the $\cal PT$-symmetric system, 
we again use 
projectors rotated relative to the optimal ones, i.e.,
\begin{equation}\label{Pi0non}
    \Pi^0(\gamma,\phi)=|\chi(\gamma,\phi)\rangle\langle\chi(\gamma,\phi)|_\eta
\end{equation}
[cf.~\eqref{nonopt1}], where the state $|\chi\rangle$ is now realized in the SLD basis in~\eqref{mu} again at some arbitrary reference point $x_0$, namely
\begin{equation}\label{chipt}
    |\chi(\gamma,\phi)\rangle=\cos\gamma\,|\mu_+(x_0)\rangle+e^{i\phi}\sin\gamma\,|\mu_-(x_0)\rangle.
\end{equation}
Similar to Example 1b, for simplicity, hereafter we set $\phi=\pi/2$.

Now, to obtain the CsFI, we need to calculate the probability of the projection measurements 
\begin{equation}\label{p1}
    p(\theta)=\langle\xi_t(\theta)|\Pi^0(\gamma)|\xi_t(\theta)\rangle_\eta.
\end{equation}
By combining Eqs.~(\ref{Fcl}), and (\ref{Pi0non})--(\ref{p1}),
one obtains
\begin{equation}\label{p0}
    p(\theta)=\dfrac{1+\cos(2\gamma)\sin2(\Omega_0t_0-\Omega t)}{2}.
\end{equation}
As a result the CsFI attains the form 
\begin{equation}\label{FclFPT}
   F_{\rm cl}^{\cal PT}(\theta,\gamma)
  =F_{\rm max}^{\cal PT}(\theta)\dfrac{\cos^22\gamma\cos^22(\Omega_0t_0-\Omega t)}{1-\cos^22\gamma\sin^22(\Omega_0t_0-\Omega t)}.
\end{equation}
Evidently, when the reference point $x_0$ is chosen to coincide with the parameter space point of the probe state $|\xi_t\rangle$, the expression in \eqref{FclFPT} simplifies to 
\begin{equation}
    F_{\rm cl}^{\cal PT}(\theta,\gamma)=F_{\rm max}^{\cal PT}(\theta)\cos^22\gamma.
\end{equation}
We illustrate the impact of reference-point choice $x_0$ for different (non-optimal) projectors on the corresponding CsFI in \figref{fig2}.

%


\newpage
\onecolumngrid

\newcommand{\appendixlabelprefix}{S:} 

\newcommand{\applabel}[1]{\label{S:#1}}
\setcounter{section}{0}
\renewcommand{\thesection}{S\arabic{section}}

\setcounter{figure}{0}
\renewcommand{\thefigure}{S\arabic{figure}}

\renewcommand{\theequation}{S\arabic{equation}}
\makeatletter
\@removefromreset{equation}{section}  
\makeatother
\setcounter{equation}{0}

\renewcommand{\appendixname}{ }

\begin{center}
    {\large\bfseries Supplementary Material for: \\
    Achieving the Quantum Fisher Information Bound in Pseudo-Hermitian Sensors, \\
    \small written by I. I. Arkhipov, F. Nori, \c{S}. K. \"Ozdemir}
\end{center}

\section{{Metric operator on a deformed Hilbert space of a pseudo-Hermitian Hamiltonian}}\label{I}
In the main text we stressed that the metric operator $\eta$ of a pseudo-Hermitian Hamiltonian \( H^{\mathrm{pH}} \) can be expressed as 
\begin{eqnarray}\label{SS}
    \eta=S^{\dagger}S,
\end{eqnarray}
where the operator $S$ determines the similarity transformations in Eq.~(1) in the main text, and which thus can be determined from the known expression of the metric.

The metric operator \( \eta \) can be straightforwardly constructed from the biorthogonal set of eigenvectors of a given  \( H^{\mathrm{pH}} \).  If \( \{|\phi_i\rangle\} \) are the left eigenvectors of \( H^{\mathrm{pH}} \), normalized such that they form a biorthonormal set with the right eigenvectors \( \{|\psi_i\rangle\} \), then the metric can be expressed as~\cite{MOSTAFAZADEH_2010}
\begin{eqnarray}
    \eta = \sum_i |\phi_i\rangle \langle \phi_i|.
\end{eqnarray}
    
This expression for $\eta$ allows to explicitly compute \( S \) by diagonalizing the former. Indeed, by decomposing the metric operator
\begin{eqnarray}\label{eta_def}
    \eta = U \Lambda U^\dagger,
\end{eqnarray}
where \( U \) is a unitary matrix whose columns are the eigenvectors of \( \eta \), and \( \Lambda \) is a diagonal matrix with positive eigenvalues of \( \eta \), a valid choice for \( S \) becomes
\begin{equation}\label{S}
    S = \sqrt{\Lambda} \, U^\dagger,
\end{equation}
such that $\eta=S^{\dagger}S$.
This relation ensures that \eqref{SS} holds, and hence defines a proper similarity transformation between the pseudo-Hermitian and Hermitian frameworks. 

As noted in the main text, the Hermitized map $S$ is not unique, and is defined up to an arbitrary unitary transformation $T$, such that two maps $S$ and $S'=TS$ give the same metric $S'^\dagger S'=S^\dagger S=\eta$. Since this unitary freedom corresponds to a gauge choice and therefore does not modify the underlying physics, one may, without loss of generality, choose 
$T$ to be parameter independent.

\section{Metric-compatible connection on a deformed Hilbert space}

As it was mentioned in the main text, the metric-compatible connection, which allows to properly define the covariant derivative of an arbitrary vector on the $\eta$-deformed Hilbert space, over a one-dimensional parameter space with variable $\theta$, has the following form:
\begin{eqnarray}\label{GS}
    \Gamma_\theta=S^{-1}\partial_\theta S.
\end{eqnarray}
Indeed, by starting from Eq.~(1) in the main text, one has
\begin{eqnarray}\label{Dtheta}
    \langle\psi^H|\partial_\theta\psi^H\rangle&=&\langle\psi^{\rm pH}|S^\dagger\partial_\theta(S|\psi^{\rm pH}\rangle) \nonumber \\
    &=&\langle\psi^{\rm pH}|S^\dagger S|\partial_\theta\psi^{\rm pH}\rangle+\langle\psi^{\rm pH}|S^\dagger \partial_\theta S|\psi^{\rm pH}\rangle \nonumber \\
    &=&\langle\psi^{\rm pH}|\eta|\partial_\theta\psi^{\rm pH}\rangle+\langle\psi^{\rm pH}|\eta (S^{-1}\partial_\theta S)|\psi^{\rm pH}\rangle \nonumber \\
    &=&\langle\psi^{\rm pH}|\eta|D_\theta\psi^{\rm pH}\rangle,
\end{eqnarray}
where we defined the covariant derivative \begin{eqnarray}
    D_\theta=\partial_\theta+S^{-1}\partial_\theta S.
\end{eqnarray}
One then immediately arrives at the general expression for the connection on the $\eta$-deformed Hilbert space $\Gamma_\theta=S^{-1}\partial_\theta S$.

Obviously, the expression for $\Gamma_\theta$ obeys the metric-compatible equation for the connection
\begin{eqnarray}
    \partial_\theta\eta-\eta\Gamma_\theta-\Gamma_\theta^{\dagger}\eta=0.
\end{eqnarray}
This  condition ensures that the norm of quantum states 
\begin{eqnarray}
    \langle\psi(\theta)|\psi(\theta)\rangle_\eta=1,
\end{eqnarray}
remains invariant under $\theta$ variation~\cite{Ju2024}.
We recall that the concept of the connection is well-known in general relativity, where one must properly define the parallel transport on the curved space~\cite{NakaharaBook}.

Note that for the case when the eigenbasis $U$  of the metric in \eqref{eta_def} is $\theta$-independent, the connection can be alternatively written solely via the metric operator as
\begin{eqnarray}\label{Geta}
    \Gamma_\theta=\frac{1}{2}\eta^{-1}\partial_\theta\eta.
\end{eqnarray}
Indeed by combining together Eqs.~(\ref{eta_def}), (\ref{S}) and (\ref{Geta}) one obtains
\begin{equation}
    \frac{1}{2}\eta^{-1}\partial_\theta\eta=\frac{1}{2}U\Lambda^{-1}(\theta)U^{\dagger}\Big( U\partial_\theta\Lambda(\theta)U^{\dagger}\Big)=\frac{1}{2}U\partial_\theta\ln\Lambda U^\dagger=U\partial_\theta\ln(\sqrt{\Lambda})U^\dagger=U\sqrt{\Lambda^{-1}(\theta)}\partial_\theta \Lambda(\theta)U^\dagger=S^{-1}\partial_\theta S.
\end{equation}
Evidently, the advantage of \eqref{Geta} when compared to \eqref{GS}, is that it enables one to straightforwardly attain the form for the connection without the need to further diagonalize the metric operator. 
Interestingly, for the two classes of pseudo-Hermitian Hamiltonians considered in the main text (namely, nonreciprocal and $\cal PT$-symmetric systems), the associated metric operators possess a 
$\theta$-independent eigenbasis $U$. As a consequence, the two definitions of the connection given in Eqs.~(\ref{GS}) and (\ref{Geta}) become equivalent.

\section{Derivation of Eq.~(5) in the main text} 
We start from Eq.~(1) in the main text, which we write down again
\begin{equation}\label{a1}
     |\psi^{\mathrm{H}}\rangle = S |\psi^{\mathrm{pH}}\rangle.
\end{equation}
Now let us explicitly calculate the standard QFI (SQFI) for the state $|\psi^{\rm H}\rangle$. 
Equation (3) in the main text, for the SQFI, can be alternatively written as
\begin{equation}\label{SQFI_def}
    F^{\rm H}(\psi^H_\theta)=4\Big(\langle\partial_\theta\psi^{\rm H}|\partial_\theta\psi^{\rm H}\rangle-|\langle\psi^{\rm H}\left |\partial_\theta\psi^{\rm H}\rangle\right |^2\Big).
\end{equation}
Let us express the first term in \eqref{SQFI_def} via the covariant derivative of the state $|\psi^{\rm pH}\rangle$:
\begin{eqnarray}
    \langle\partial_\theta\psi^{\rm H}|\partial_\theta\psi^{\rm H}\rangle&&=\Big(\langle\partial_\theta\psi^{\rm pH}|S^{\dagger}+\langle\psi^{\rm pH}|\partial_\theta S^\dagger\Big)\Big( S|\partial_\theta\psi^{\rm pH}\rangle+\partial_\theta S|\psi^{\rm pH}\rangle\Big) \nonumber \\
    &&=\langle\partial_\theta\psi^{\rm pH}|S^{\dagger}S|\partial_\theta\psi^{\rm pH}\rangle+\langle\partial_\theta\psi^{\rm pH}|(S^{\dagger}S)(S^{-1}\partial_\theta S)|\psi^{\rm pH}\rangle \nonumber \\
    &&\ \ \ \ + \langle\psi^{\rm pH}|(\partial_\theta S^{\dagger}S^{\dagger -1})(S^{\dagger}S)|\partial_\theta\psi^{\rm pH}\rangle+
    \langle\psi^{\rm pH}|(\partial_\theta S^{\dagger}S^{\dagger -1})(S^{\dagger}S)(S^{-1}\partial_\theta S)|\partial_\theta\psi^{\rm pH}\rangle \nonumber \\
    && = \langle D_\theta\psi^{\rm pH}|\eta|D_\theta\psi^{\rm pH}\rangle,
\end{eqnarray}
where as before the metric operator $\eta=S^\dagger S$.
Similarly, from \eqref{Dtheta}, one has $\langle\psi^{\rm H}|\partial_\theta\psi^{\rm H}\rangle=\langle \psi^{\rm pH}|\eta|D_\theta\psi^{\rm pH}\rangle$.
As a result one arrives at Eq.~(5) in the main text, namely
\begin{equation}
    F^{\rm H}(\psi^H_\theta)=F^{\rm pH}_c(\PV_{\theta}).
\end{equation}

\section{Derivation of the metric operator for the pseudo-Hermitian Hamiltonian analyzed in Example 1.b in the main text}
We begin our analysis with a general $2 \times 2$ real-valued nonreciprocal pseudo-Hermitian Hamiltonian of the form
\begin{equation}\label{H1}
\PH = \begin{pmatrix}
\omega & k_1 \\
k_2 & -\omega
\end{pmatrix}, \quad k_1 \neq k_2, \quad k_{1,2}>0.
\end{equation}
Finding the metric for this general form would allow us to immediately determine the metrics for Hamiltonians analyzed in both {\it Example 1.a} and {\it Example 1.b}.

First, we find the associated Hermitian Hamiltonian $\HH$, and the corresponding similarity transformations $S$ with metric $\eta$. 
The transformation $S$ is known in the literature on the non-Hermitian skin effect~\cite{arkhipov2023a}, and reads as follows
\begin{eqnarray}
     S={\rm diag}\left[1,\sqrt{k_1/k_2}\right],
\end{eqnarray}
which gives the associated Hermitian Hamiltonian $\HH=S\PH S^{-1}$ as follows
\begin{equation}\label{Hh1}
\HH = \begin{pmatrix}
\omega & \sqrt{k_1k_2} \\
\sqrt{k_1k_2} & -\omega
\end{pmatrix},
\end{equation}
The explicit form for $S$ immediately gives a metric 
\begin{eqnarray}\label{etanon}
    \eta=S^{\dagger}S={\rm diag}\left[1,k_1/k_2\right].
\end{eqnarray}
implying that the metric is already in the diagonal form, and therefore its eigenbasis $U=I_2$ is always independent of any system parameter.

From \eqref{etanon} one immediately obtains the form of the metric operators defined on the $\eta$-modified Hilbert space generated by non-Hermitian Hamiltonians in Eqs.~(9) and (11), namely
\begin{eqnarray}
    \eta=\begin{pmatrix}
        1 & 0 \\
        0 & \delta^{-2}
    \end{pmatrix}, 
 \end{eqnarray}
 and
 
\begin{eqnarray}\label{etanonex}
 \eta=\begin{pmatrix}
        1 & 0 \\
        0 & \dfrac{\delta^{-1}+\theta}{\delta+\theta}
    \end{pmatrix},
\end{eqnarray}
respectively.

\begin{figure}[t!]
    \includegraphics[width=0.77\textwidth]{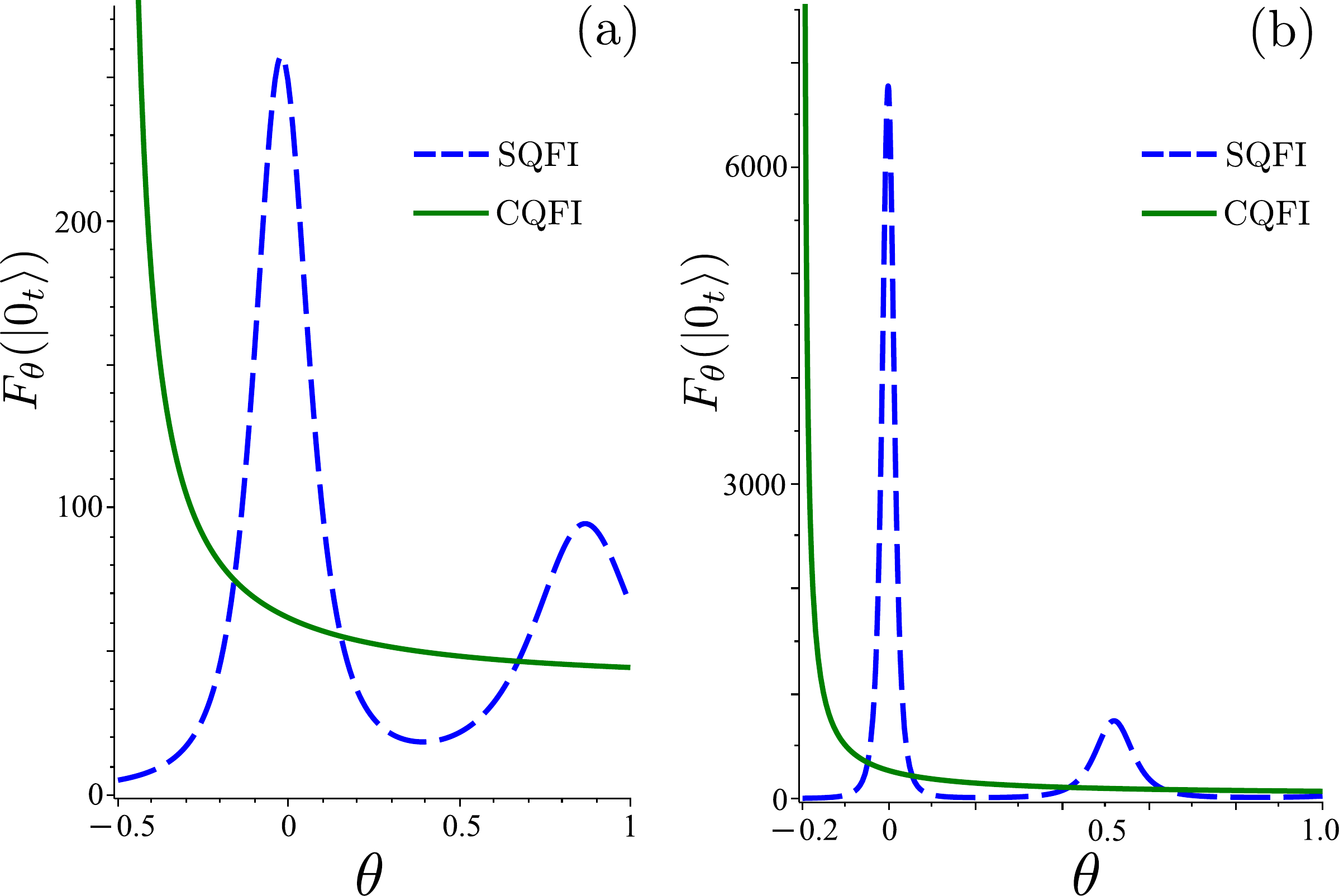}
    \caption{QFI for the state $|0_t\rangle$ in \eqref{0t}, calculated using the SQFI (blue dashed curves) and CQFI (green solid curves) formalisms. The dynamics is governed by the pseudo-Hermitian Hamiltonian in \eqref{Hex}. The system parameters are: (a) $\delta=0.5$ [arb. units], and (b) $\delta=0.2$ [arb. units]. The time is set to $t=\pi$ [arb. units]. The EPs are defined by $\theta_{\rm EP}=-0.5$ [arb. units] on panel (a), and $\theta_{\rm EP}=-0.2$ [arb. units] on panel (b). The locations of the EPs $\theta_{\rm EP}$ coincide with the positions of the $y$-axis on both panels.}
    \label{fig}
\end{figure}

\section{Comparison between covariant and standard QFI calculated for states described by the nonreciprocal pseudo-Hermitian Hamiltonian given in Example 1.b in the main text}

Here, we briefly discuss the implications of applying the {\it standard} QFI formalism, rather than the correct {\it covariant} QFI, to quantum systems governed by pseudo-Hermitian Hamiltonians. We explicitly demonstrate  that such an approach in general  leads to misleading conclusions. For simplicity, we focus on a linearly perturbed nonreciprocal pseudo-Hermitian Hamiltonian analyzed in Example 1.b and given in Eq.~(11) in the main text, namely:
\begin{equation}\label{Hex}
    \PH(\theta)=\begin{pmatrix}
        0& \delta^{-1}+\theta \\
        \delta+\theta & 0
    \end{pmatrix}, \quad \delta\neq 0,
\end{equation} 

We initialize the system in the vacuum state $|0\rangle=[1,0]^T$ and  calculate both the SQFI and CQFI, using Eqs.~(3) and (4), respectively, for the time-evolving state $|0_t\rangle\equiv\exp(-i\PH t)|0\rangle$, which is given by Eq.~(13) in the main text, namely
\begin{equation}\label{0t}
    |0_t\rangle\equiv\cos(\sqrt{k_1k_2}t)|0\rangle-i\sqrt{k_2/k_1}\sin(\sqrt{k_1k_2}t)|1\rangle,
\end{equation}
where $k_1=\delta^{-1}+\theta$, and $k_2=\delta+\theta$. 

On the flat Hilbert space with $\eta=I$, the evolving state $|0_t\rangle=|0_t\rangle/\sqrt{N}$ has to be renormalized at each time $t$, since
\begin{eqnarray}
    N(t)_{\rm flat}=\cos^2(\sqrt{k_1k_2}t)+\sin^2(\sqrt{k_1k_2}t)k_2/k_1.
\end{eqnarray} 

On the contrary, within the deformed Hilbert space formalism, the normalization for the same evolving state is automatically ensured by the corresponding metric $\eta$, i.e., $\langle 0_t|\eta|0_t\rangle=1$. Specifically, here the normalization constant does not depend on time and equals $N_{\rm deform}=1$. 

We plot both the SQFI and CQFI for the state $|0_t\rangle$ in \figref{fig}. As illustrated on the graph, by naively applying the SQFI formalism to the state $|0_t\rangle$, one completely misses the onset of the spectral phase transition occurring at the EP. 

On the contrary, the CQFI framework faithfully captures the {\it phase transition at the EP}, associated with the divergent behavior of the $F^{\rm pH}_c(|0_t\rangle)$. This demonstrates that the CQFI correctly reflects the system's sensitivity enhancement near the EP; thus making it a powerful tool for the proper analysis of parameter estimation in pseudo-Hermitian systems.

\section{Calculation of the eigenbasis of the covariant symmetric logarithmic derivative operator in Eq.~(14) in the main text}
The explicit form for the covariant symmetric logarithmic derivative (CSLD) operator $L_c$ determined by the time-evolving state $|\psi_t\rangle$ in Eq.~(13) in the main text reads as
\begin{eqnarray}\label{Lc}
    L_c=2(|D_\theta\psi_t\rangle\langle\psi_t|_\eta+|\psi_t\rangle\langle D_\theta\psi_t|_\eta),
\end{eqnarray}
and where the metric operator $\eta$ is given in \eqref{etanonex}. The derivation of the CSLD eigenbasis for the $\cal PT$-symmetric system proceeds in an analogous manner.

First, we explicitly calculate the state $|D_\theta\psi_t\rangle$.
Exploiting the fact that the metric eigenbasis is $\theta$-independent, according to~\eqref{etanon}, one can use \eqref{Geta} for the connection $\Gamma_\theta$.

The inverse of $\eta$ is
\begin{eqnarray}
    \eta^{-1}=\begin{pmatrix}
        1 & 0 \\
        0 & \dfrac{\delta+\theta}{\delta^{-1}+\theta}
    \end{pmatrix}.
\end{eqnarray}
The derivative $\partial_\theta\eta$ then reads
\begin{eqnarray}
    \partial_\theta\eta=\begin{pmatrix}
        0 & 0 \\
        0 & \dfrac{\delta-\delta^{-1}}{(\delta+\theta)^2}
    \end{pmatrix}.
\end{eqnarray}
Thus the connection can be written as
\begin{eqnarray}
    \Gamma_\theta=\frac{1}{2}\eta^{-1}\partial_\theta\eta=\begin{pmatrix}
        0 & 0 \\
        0 & \dfrac{\delta-\delta^{-1}}{2(\delta+\theta)(\delta^{-1}+\theta)}
    \end{pmatrix}.
\end{eqnarray}
This means that for an arbitrary vector $v=(x_1,x_2)^T$, the covariant derivative of the vector attains the form
\begin{eqnarray}\label{Dv}
    D_\theta v = \left(\partial_\theta x_1,\partial_\theta x_2+\dfrac{\delta-\delta^{-1}}{2(\delta+\theta)(\delta^{-1}+\theta)}x_2\right)^T.
\end{eqnarray}
Now recalling the form of the state $|\psi_t\rangle$ in Eq.~(13), which for completeness we write here as
\begin{eqnarray}\label{psi}
    |\psi_t\rangle = \cos(\sqrt{k_1k_2}\,t)\,|0\rangle 
    - i\sqrt{\frac{k_2}{k_1}}\,\sin(\sqrt{k_1k_2}\,t)\,|1\rangle,
\end{eqnarray}
then, with the help of \eqref{Dv}, one finds its covariant derivative as follows
\begin{eqnarray}\label{Dpsi}
    |D_\theta\psi_t\rangle=-\dfrac{t(k_1+k_2)}{2\sqrt{k_1k_2}}\left[\sin(\sqrt{k_1k_2}t)|0\rangle+i\sqrt{\dfrac{k_2}{k_1}}\cos(\sqrt{k_1k_2})|1\rangle\right],
\end{eqnarray}
where $k_1=\delta^{-1}+\theta$, and $k_2=\delta+\theta$.

A straightforward calculation shows that the states $|\psi_t\rangle$ and $|D_\theta\psi_t\rangle$ are $\eta$-orthogonal to each other, i.e., $\langle\psi_t|\eta|D_\theta\psi_t\rangle=0$.
This implies that the operator $L_c$ acts nontrivially on the space spanned by $\{|\psi_t\rangle,|D_\theta\psi_t\rangle\}$, whose superpositions can thus define the eigenstates we are after. 
We denote the $\eta$-normalized state $|D_\theta\psi_t\rangle$ as $|u_t\rangle$, which reads
\begin{eqnarray}
    |u_t\rangle = \sin(\sqrt{k_1k_2}\,t)\,|0\rangle 
    + i\sqrt{{k_2}/{k_1}}\,\cos(\sqrt{k_1k_2}\,t)\,|1\rangle.
\end{eqnarray}
One then verifies that the combinations 
\begin{eqnarray}\label{opt}
    |\nu_\pm\rangle=(|\psi_t\rangle\pm|u_t\rangle)/\sqrt{2}, \quad \langle\nu_i|\nu_j\rangle_\eta=\delta_{ij}, \quad i,j=\pm,
\end{eqnarray}
constitute the eigenbasis of the covariant SLD operator in \eqref{Lc} with real eigenvalues
\begin{eqnarray}
    \lambda_\pm=\pm\sqrt{F^{\rm pH}_{\rm max}(\theta)},
\end{eqnarray}
being equal (with plus-minus sign) to the square root of the upper bound of the covariant quantum Fisher information given in Eq.~(12) in the main text, whose form is
\begin{eqnarray}
    F^{\rm pH}_{\rm max}(\theta)= \frac{(\delta^2+2\theta\delta+1)^2}{\delta(\delta+\theta)(\delta\theta+1)}t^2.
\end{eqnarray}
Consequently, by preparing the given system into the state $|\psi_t\rangle$ and by performing projections on the eigenbasis of $L_c$ one can achieve the ultimate precision in the parameter estimation.

\section{Calculation of covariant classical Fisher information from non-optimal projections}\label{VIII}
In {\it Example 1.b} of the main text, we presented the final expression for the covariant classical Fisher information (CsFI) obtained when the system's state is projected onto non-optimal bases, constructed as rotated eigenstates of the covariant SLD operator. Here, we provide its explicit calculation. 

In Eq.~(15) in the main text we introduced the nonoptimal projectors, namely
\begin{equation}\label{nonopt}
  \Pi(\gamma,\phi)=|\chi(\gamma,\phi)\rangle\langle\chi(\gamma,\phi)|_\eta, \quad \text{and} \quad {\mathds 1}-\Pi(\gamma,\phi),
\end{equation}
where
\begin{eqnarray}
    |\chi(\gamma,\phi)\rangle=\cos\gamma\,|\nu_+\rangle+e^{i\phi}\sin\gamma\,|\nu_-\rangle,
\end{eqnarray}
denotes a generic state rotated away from the optimal basis in~\eqref{opt}. In the following we assume $\phi=\pi/2$.

As noted in the main text, the reference point in parameter space for defining the eigenbasis of the covariant SLD operator can be chosen arbitrarily, since projections onto this basis at any point saturate the CsFI to the CQFI bound; a statement we further confirm below. 
We denote these state as follows
\begin{eqnarray}\label{nu0}
    |\nu_{\pm}^0\rangle=(\cos\xi_0\pm\sin\xi_0)|0\rangle- is_0(\sin\xi_0\mp\cos\xi_0)|1\rangle,
\end{eqnarray}
where $\xi_0=\sqrt{k_1^0k_2^0}t_0$, $s_0=\sqrt{k_2^0/k_1^0}$, and $k_{1,2}^0$ are defined as in \eqref{psi} by some arbitrary  $\delta_0$ and $\theta_0$. 
As a result, we obtain the non-optimal projector $\Pi^0(\gamma)=|\chi^0(\gamma)\rangle\langle\chi^0(\gamma)|_{\eta_0}$ and its complement in \eqref{nonopt}.

Now, to obtain the CsFI, we need to calculate the probability of the projection measurements 
\begin{eqnarray}\label{p}
    p(\theta)=\langle\psi_t(\theta)|\Pi^0(\gamma)|\psi_t(\theta)\rangle_\eta.
\end{eqnarray}
Before proceeding, we note an important issue concerning the definition of the metric in probability calculations. In general, the states $|\nu(x_0)\rangle$ and $|\psi_t\rangle$ correspond to distinct local metrics, which makes their inner product ambiguous. This ambiguity can be resolved either by parallel transporting one state to the location of the other, where they can be compared, or by flattening the $\eta$-deformed Hilbert space. Here we adopt the latter approach.

Specifically, the inner product $\langle\psi_t|\nu_\pm^0\rangle_\eta$ becomes well defined once the $\eta$-deformed space is flattened, as described in Eq.~(1) of the main text. For the pseudo-Hermitian Hamiltonians considered here, states on the flat Hilbert space and those on the $\eta$-modified space are connected by a similarity transformation, $|\psi^H\rangle=S|\psi^{\rm pH}\rangle$. Accordingly, the inner product $\langle\psi_t|\nu_\pm^0\rangle_\eta$ has a well-defined counterpart on the flat space,
\begin{eqnarray}\label{etap}
    \langle\psi_t^H|\nu_\pm^{0H}\rangle=\langle\psi_t|S^{\dagger}_\psi S_{\nu_\pm^0}|\nu_\pm^0\rangle,
\end{eqnarray}
which identifies the proper metric operator as $\eta_p=S^\dagger_\psi S_{\nu_\pm^0}$. The matrices $S_\psi$ and $S_{\nu^0}$ are obtained directly from the corresponding local metrics, according to \eqref{S}.

Now, by combining Eqs.~(\ref{nonopt})--(\ref{etap}), one obtains
\begin{eqnarray}\label{pth}
    p(\theta)=\dfrac{1}{2}\left[1+\cos(2\gamma)\sin2(\xi_0-\sqrt{k_1k_2}t)\right].
\end{eqnarray}
Clearly, as it follows from \eqref{pth}, one verifies $0\leq p(\theta)\leq 1$.
Accordingly, the probability measuring the complement projector ${\mathds 1}-\Pi(\gamma,\phi)$ is $1-p(\theta)$.
\begin{figure}[t!]
    \includegraphics[width=0.65\textwidth]{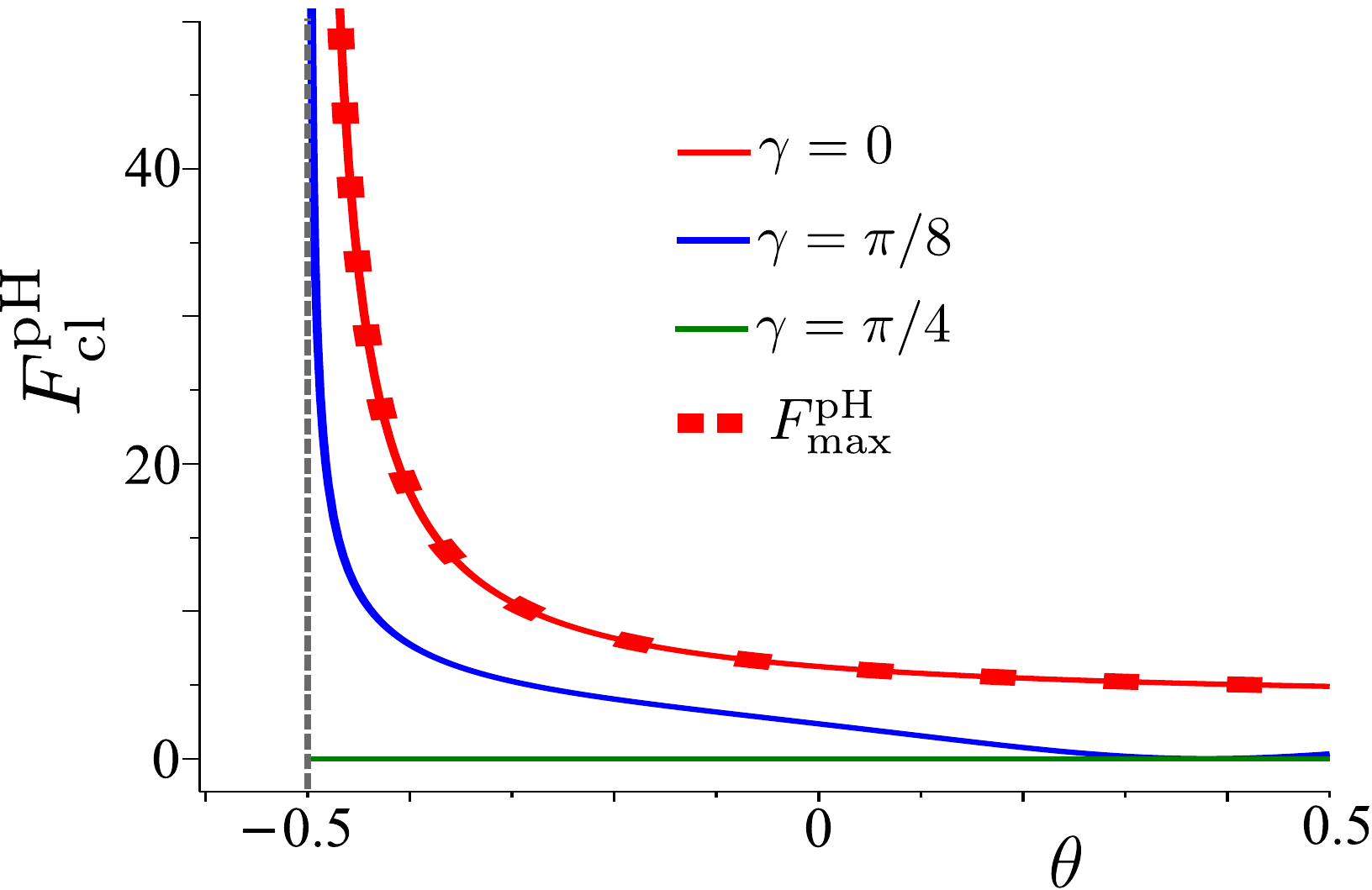}
    \caption{This figure is a complement to Fig.~1(a) in the main text for the case when the reference point $x_0$, chosen for the projectors, does not coincide with that of the system's state $|\psi_t\rangle$. The chosen system parameters are: (a) $\delta_0=\delta=1/2$, $\theta_0=-1/4$, $t_0=t=1$. 
    }
    \label{S2}
\end{figure}

Now by recalling the definition of the CsFI (given in Eq.~(8) in the main text), and applying it to our studied case here we get
\begin{eqnarray}\label{Fclsup}
    F_{\rm cl}^{\rm pH}(\theta)=\sum{(\partial_\theta p_i)^2}/{p_i}=\dfrac{\left[\partial_\theta p(\theta)\right]^2}{p(\theta)[1-p(\theta)]}.
\end{eqnarray}
By inserting the expression for $p(\theta)$, given in~\eqref{pth}, into \eqref{Fclsup}, we finally attain
\begin{eqnarray}\label{Fcl1}
    F_{\rm cl}^{\rm pH}(\theta)=F^{\rm pH}_{\rm max}(\theta)\dfrac{\cos^22\gamma\cos^22(\xi_0-\sqrt{k_1k_2}t)}{1-\cos^22\gamma\sin^22(\xi_0-\sqrt{k_1k_2}t)}\leq F^{\rm pH}_{\rm max}(\theta),
\end{eqnarray}
where the upper bound of the CQFI is given in Eq.~(12) in the main text, or more explicitly
\begin{eqnarray}
    F^{\rm pH}_{\rm max}(\theta)=\frac{(\delta^2+2\theta\delta+1)^2}{\delta(\delta+\theta)(\delta\theta+1)}t^2.
\end{eqnarray}

Assuming that  $\xi_0=\sqrt{k_1k_2}t$, i.e., the chosen reference point coincides with the point in parameter space for the measured state $|\psi_t\rangle$, \eqref{Fcl1} simplifies to 
\begin{eqnarray}\label{Fcl3}
    F_{\rm cl}^{\rm pH}(\theta)=F^{\rm pH}_{\rm max}(\theta)\cos^22\gamma.
\end{eqnarray}
In the main text in Fig.~1, we presented the plots for the CsFI for this particular case given in \eqref{Fcl3}. As a complement to Fig.~1, we illustrate the behavior of the CsFI when the reference point $x_0$ does not coincide with the parameter space point of the measured state $|\psi_t\rangle$ in Fig.~\ref{S2}.

As expected, the maximum for CsFI, which is saturated to the CQFI bound, can only be achieved when the projection states $|\chi\rangle$ are those of the covariant SLD operator, for which $\gamma=n\pi$, $n\in{\mathbb Z}$. Otherwise, the CsFI is always suppressed, reaching zero values when $\gamma=(2k+1)\pi/4$, $k\in{\mathbb Z}$.

\section{Universality of the covariant symmetric logarithmic derivative eigenbasis}
In the main text and at the beginning of Sec.~\ref{VIII} of this Supplementary Material we emphasized that the eigenbasis of the covariant  symmetric logarithmic derivative operator is universal with respect to the choice of reference point in parameter space. In other words, projecting the system’s state onto this eigenbasis at any reference point saturates the CsFI to the upper bound of the CQFI. While the individual projection probabilities depend on the chosen reference point, the resulting CsFI remains invariant.

From \eqref{Fcl1} it is now evident why it is so. Indeed, by setting $\gamma=0$, the nominator and denominator in the fraction in the r.h.s. of \eqref{Fcl1} become equal. In other words, for any $\xi_0$ one has
\begin{eqnarray}
    F_{\rm cl}^{\rm pH}(\theta)=F^{\rm pH}_{\rm max}(\theta).
\end{eqnarray}

\end{document}